\documentclass[aps,twocolumn,showpacs,superscriptaddress,floatfix,letter]{revtex4}
\usepackage{graphicx}
\usepackage{psfrag}
\usepackage{color}
\usepackage{float}
\usepackage{braket}
\usepackage{amsmath}     
\usepackage{amsthm}
\usepackage{amsfonts}
\usepackage{amssymb}
\usepackage{mathrsfs}
\usepackage{fancyhdr}
\usepackage{hyperref}
\usepackage{bm}

\begin{document}
\newcommand{\be}{\begin{equation}}
\newcommand{\ee}{\end{equation}}
\newcommand{\ben}{\begin{eqnarray}}
\newcommand{\een}{\end{eqnarray}}
\newcommand{\baq}{\begin{array}}
\newcommand{\eaq}{\end{array}}
\newcommand{\om}{(\omega )}
\newcommand{\bef}{\begin{}}
\newcommand{\eef}{\end{figure}}
\newcommand{\leg}[1]{\caption{\protect\rm{\protect\footnotesize{#1}}}}
\newcommand{\ew}[1]{\langle{#1}\rangle}
\newcommand{\no}{\nonumber}
\newcommand{\geff}{g_{\mbox{\it{\scriptsize{eff}}}}}
\newcommand{\da}[1]{{#1}^\dagger}
\newcommand{\cf}{{\it cf.\/}\ }
\newcommand{\ie}{{\it i.e.\/}\ }
\newcommand{\he}{\mathcal{H}}
\setlength\abovedisplayskip{5pt}
\setlength\belowdisplayskip{5pt}

\title{Cooperative robustness to dephasing: single-exciton Superradiance
in a nanoscale ring to model natural light-harvesting systems}

\author{G.~Luca \surname{Celardo}}
\affiliation{Dipartimento di Matematica e
Fisica, Universit\`a Cattolica del Sacro Cuore, via Musei 41, I-25121 Brescia, Italy
}
\affiliation{Interdisciplinary Laboratories for Advanced Materials Physics,
 Universit\`a Cattolica del Sacro Cuore, via Musei 41, I-25121 Brescia, Italy}
\affiliation{Istituto Nazionale di Fisica Nucleare, Sezione di Pavia, 
via Bassi 6, I-27100, Pavia, Italy}
\author{Paolo \surname{Poli}}
\affiliation{Dipartimento di Matematica e
Fisica, Universit\`a Cattolica del Sacro Cuore, via Musei 41, I-25121 Brescia, Italy
}
\author{Luca \surname{Lussardi}}
\affiliation{Dipartimento di Matematica e
Fisica, Universit\`a Cattolica del Sacro Cuore, via Musei 41, I-25121 Brescia, Italy
}
\author{Fausto \surname{Borgonovi}}
\affiliation{Dipartimento di Matematica e
Fisica, Universit\`a Cattolica del Sacro Cuore, via Musei 41, I-25121 Brescia, Italy
}
\affiliation{Interdisciplinary Laboratories for Advanced Materials Physics,
 Universit\`a Cattolica del Sacro Cuore, via Musei 41, I-25121 Brescia, Italy}
\affiliation{Istituto Nazionale di Fisica Nucleare, Sezione di Pavia, 
via Bassi 6, I-27100, Pavia, Italy}

\begin{abstract}  
We analyze a 1-d ring structure composed of many
two-levels systems, in the limit where only one excitation is present.
The two-levels systems are coupled to a common environment, where the
excitation can be lost,
which induces super and subradiant behavior. 
Moreover, each two-levels system is coupled to another independent
environment, modeled by a classical white
noise, simulating a dephasing bath 
and described by the Haken-Strobl master equation. 
Single exciton Superradiance,  an example of cooperative quantum coherent
effect, is 
destroyed at a critical dephasing strength proportional to the system size, showing robustness
of cooperativity to the action of the dephasing environment.
We also show that the 
coupling to a common decay channel 
 contrasts the action of 
dephasing, driving the entanglement decay to slow down on increasing the system size. 
Moreover, after  a projective measurement 
which finds the excitation in the system, the entanglement reaches a stationary value, 
independent of the initial conditions.  
\end{abstract}                                                               
                                                                            
\date{\today}
\pacs{71.35.-y, 03.65.Yz, 05.60.Gg}
\maketitle

\section{Introduction}

Emergent properties arise due to cooperative behavior of the many
constituents of a complex system. They belong  to the system as a whole and
not to its constituents, and 
are at the center of many research fields in condensed matter physics. 
In quantum systems, differently from classical ones, 
additional emergent properties are possible due to quantum coherence.
Among the many fascinating aspects of these properties, one important open question
regards their robustness  to the effects induced by the presence of an environment. 
This robustness might enable to exploit coherent quantum effects to
build quantum devices for 
information technologies and basic energy science.  

As an example of quantum coherent emergent property, we consider here 
single exciton Superradiance~\cite{scully}. 
Superradiance was originally discovered in the context of atomic
clouds interacting with the electromagnetic field
(EMF), see the seminal paper by Dicke~\cite{dicke54}, where 
Superradiance 
involved many excited atoms (each modeled as a two-levels system). 
Dicke showed that, when 
the wavelength associated with the emitted photon is larger than
the system size,
there are some
states which emit light with an intensity proportional to $N^2$,
where $N$ is the number of atoms. This kind of Superradiance would
occur also in an ensemble of classical emitters with proper initial
conditions.
However, Superradiance can occur also when only one
excitation is present in the system. In this case it
 is a purely quantum effect~\cite{scully}, due
to exciton states extended over many sites (representing a fully
entangled state of the two-levels systems sharing the excitation). 
The  decay rate of such a state is
cooperatively enhanced w.r.t.~the single system decay rate, in that it is proportional to 
the system size.
For instance, in molecular aggregates~\cite{mukameldeph}, and light
harvesting complexes~\cite{schulten}, the single excitation
approximation is valid under natural light intensity since solar light
is very dilute. For molecular aggregate
if we call $\gamma$ the radiative decay
width of an isolated molecule, an excitation spread coherently over $N$
molecules has a radiative decay width equal to $N \gamma$. 
Note that Superradiance comes always together with Subradiance, 
that is the existence of other states with a  suppressed decay width 
(\emph{i.e.}\ smaller than the single system decay width).

Superradiant behavior 
with respect to the EMF was found experimentally 
in many molecular aggregates~\cite{Jaggr,vangrondelle}. 
Superradiance does not occur only w.r.t. the EMF, as it is commonly
considered in literature, but  it is a  general phenomenon in open quantum
systems~\cite{Zannals} which can occur whenever many discrete quantum levels are
coupled to some common decay channels, characterized by a continuum of states. 
In particular, its relevance in quantum transport, where the decay
channels represent the continuum of scattering states, was pointed out in
Ref.~\cite{kaplan}. 
Since the discovery of  quantum coherent effects 
in biological systems even at room temperature~\cite{photo,photoT,photo2,photo3,schulten}, Superradiance has been thought
to have a functional role in photosynthesis,  w.r.t.~both the
absorption of electromagnetic radiation and the transfer of the excitation
to another absorbing complex, such as the reaction
center~\cite{schulten,srlloyd,sr2,srfmo,srrc}.
In the latter case, 
the same effects found for the electromagnetic field appear,
for instance we can have
Sub and Superradiance in transport which implies that the excitation can be transferred
to the reaction center or to a lead with an enhanced or suppressed rate.
Moreover, proposals of solid state quantum devices 
for photon sensing and light harvesting
based on Superradiance have been made~\cite{superabsorb}.

These systems  are usually  subject to the effects of different
environments. Together with a common decay channel,  
where the excitation can be
lost either by recombination and photon emission or by trapping to a
central core absorber, there are also 
other  environments (such as a phonon bath) which 
induce various kinds of noise characterized by different correlation time-scales, to be compared to the excitonic transport time: i) short-time correlations
giving rise to  dephasing (homogeneous broadening), and ii) long-time correlations producing
on-site static disorder (inhomogeneous broadening).
The interplay of different environments is essential to determine the
efficiency of light absorption and excitation transfer.

The interplay of  dephasing 
and Superradiance  has been considered in many publications,
for instance in Ref.~\cite{pra} the case of many atoms interacting with
the EMF in presence of collisional dephasing has been
analyzed  for many excitations.
On the contrary, here, we focus our attention on 
the relation between dephasing and Superradiance where only one excitation
is present in the system. This case turns out to
be  more relevant for light harvesting complexes, as explained above.

The interplay of single excitation Superradiance and dephasing has been analyzed in Ref.~\cite{mukameldeph}
giving more attention to the behavior of populations
and not  to coherences. Specifically, our main interest 
is addressing  the question of  
 how cooperativity can protect 
coherences and entanglement.

The coupling to a common decay channel, which
induces Superradiance, is taken into
account by means of an effective non-Hermitian
Hamiltonian~\cite{Zannals,rotterb,alberto}.
Within this framework, it is possible to recover 
the generation of entanglement found in the case of two qubits coupled to a common
decay channel in Ref.~\cite{plastina}.

The coupling to a dephasing environment 
is modeled as stochastic fluctuations of diagonal energies,
a common way to include dephasing in exciton dynamics, and it will
be taken into account by using the Haken--Strobl approach~\cite{HS}.
The Haken--Strobl approach
has been widely used in the
past to include  dephasing~\cite{mukameldeph,expdeph}
and it has also  been analyzed in many recent applications
~\cite{lloyd,deph,cao} for 
its simplicity and  effectiveness in describing strong
dephasing in the high-temperature limit.
Despite the  essential features
captured by this exactly solvable model,
concerning the interplay between coherent dynamics and dephasing,
its limitations in describing finite temperature experimental data
are well known ~\cite{spano}. 
Indeed it  represents the case of strong collisions between
excitons and phonons (infinite temperature) thus overestimating the real
effect of dephasing on coherences. 
Even if one may expect that
the dephasing induced by stochastic fluctuations hinders  quantum coherent effects,
 and thus Superradiance,  we will see that the presence of a 
common decay channel 
is able to partially limit the effect of such a noise.

In this paper we will not consider the effect of a static disorder
(inhomogeneous broadening),
which has been analyzed by some of the Authors of this paper~\cite{laltro}.
The interplay of dephasing (homogeneous broadening) and static disorder
has also been analyzed in Refs.~\cite{kos,cao}.

\section{The Model} 

We consider here a paradigmatic model with $N$ two-levels systems  arranged in a ring structure, 
which has been also analyzed  
in several papers~\cite{cao,superabsorb,sarovarbio,fassioli,mukameldeph,mukamelspano} 
to describe different systems, such as 
molecular  J-aggregates~\cite{Jaggr},
 bio-inspired devices for photon sensing~\cite{superabsorb},
and efficient light-harvesting systems~\cite{sarovarbio}.
In particular, it has been often considered in the frame of exciton transport in 
natural photosynthetic complexes, such as LHI, LHII,
where chlorophyll molecules 
are held in a ring-like structure by a protein scaffold~\cite{schulten1}.
\begin{figure}[t]
\vspace{0cm}
\includegraphics[width=8cm,angle=0]{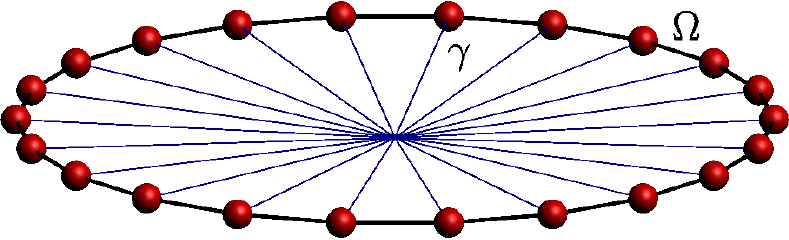}
\caption{ (Color online)
$N$ sites with  coupling $\Omega$ connected to a common
decay channel with the same  coupling strength $\gamma$.
}
\label{ring}
\end{figure}

Usually, under low light intensity,  only one excitation is considered
and the system becomes
equivalent to a tight binding model where the excitation  can hop from
site to site, as described in Fig.~\ref{ring}. 
Specifically, we consider a ring with nearest neighbors coupling, $\Omega$, 
 described by the following periodic tight binding Hamiltonian:
\be
\label{ham1}
H^{tb} = \sum_{j=1}^N E_j \ket{j}\bra{j} \ + \\
 \Omega \sum_{j=1}^{N}\left(\ket{j}\bra{j+1} + \ket{j+1}\bra{j}\right). 
\ee
We also fix the energy scale taking $\Omega=1$  and, for simplicity,
$E_{j}=0$, for  $j=1,\ldots ,N$.
Here  $|j\rangle$ represents  a state in which the excitation is at the site $j$,
while all the other sites are unoccupied. In terms of two-level system
states ($|0\rangle, 
|1\rangle$) it can be written  as
$$ |j\rangle = |0\rangle_1 |0\rangle_2 \ldots |1\rangle_j \ldots |0\rangle_N\,. $$

Each site is coupled to a common decay channel
with coupling strength $\gamma$, 
as shown in Fig.~\ref{ring}. 
\be
\label{ham2}
\he = H^{tb}-i\frac{\gamma }{2} Q , \qquad
 Q_{ij}=1,  \qquad
 i,j=1,\ldots,N.
\ee
 In the case of molecular aggregate this common decay channel can represent 
 the coupling with the EMF~\cite{mukameldeph} where the excitation can be lost.
Indeed the non-Hermitian Hamiltonian used to model the coupling of
molecular aggragates with the EMF~\cite{mukameldeph}
reduces to Eq.(\ref{ham2}), in the limit of large wavelength with 
respect to the system size and for  molecular aggregates with
parallel dipoles.  For instance the wavelength of the
absorbed light (hundreds of nanometers) is much larger than the system
size of natural complexes 
such as LHI, LHII (few tens of nanometers)~\cite{schulten}.

We stress that while in the case of  the EMF, $\gamma$ and $\Omega$ have a common origin,
since they both arise from the interaction  with the EMF, 
 here we take a more general point of view:
$\gamma$ can also represent the coupling with a different
environment, so that it  can be varied independently of $\Omega$.
For instance in molecular aggregates $\gamma$ can also represent  the
coupling to a central core absorber, 
such as  the reaction center~\cite{srfmo,srrc} or a semi-infinite lead
where the excitation 
can be trapped ~\cite{inprep,superabsorb,kaplan,rotterb}.

Due to its specific structure, the operator $Q$ in Eq.(\ref{ham2}) has only one eigenstate
with a non-zero eigenvalue:  this is the fully extended state
with eigenvalue equal to $N$,
\begin{equation}
|S \rangle = \frac{1}{\sqrt{N}}    \sum_{i=1}^N  |i\rangle.
\label{Sr}
\end{equation}
This eigenstate also corresponds to the
ground state of $H^{tb}$.
All the other eigenstates of $Q$ are degenerate with null eigenvalues and, since $[Q, H^{tb}] = 0$,
they can be chosen to match the other eigenstates of $H^{tb}$, see Ref.~\cite{laltro}.
 This implies that only the state
$|S \rangle$, Eq.~(\ref{Sr}), has a non-vanishing decay width
equal to the total decay width of the system:
$\Gamma_N=N\gamma$. This is the superradiant state and
 the  dependence on $N$  of the decay width is the
hallmark of the cooperative nature of Superradiance.
All the other states with zero decay width
are called subradiant. Clearly, this  perfect segregation of the decay
widths will not persists in presence of dephasing or of any other
source of disorder which breaks the symmetry of the model.

Thus in our model, in absence of disorder, we are in a superradiant
regime for any $\gamma$ and we have: 
\begin{itemize}
\item a superradiant state,
$|S \rangle$
with a decay width given by $N \gamma$,
\item  $N-1$ subradiant states, orthogonal
to the superradiant one, with zero  decay width for any value of $\gamma$. 
\end{itemize}
Note that this situation is a peculiarity of the model:  usually, the
superradiant regime sets in only above  a critical value of the
coupling with the continuum of states. This fact shows that ring-shaped nanostructures are ideal to exploit Superradiance, and it might also explain why they appear 
 in natural
photosynthetic complexes.

The effect of the dephasing environment is taken into account in the
frame of the Haken-Strobl (HS) model~\cite{HS}. 
In the HS model considered below, dephasing arises due to uncorrelated
fluctuations of site energies,
\begin{equation}
\label{flu}
\langle E_i(t) E_j(t') \rangle = \hbar \delta_{ij} \delta(t-t') \gamma_{\phi}.
\end{equation}
where $\gamma_\phi$ is the dephasing rate measured in units of energy, proportional to
the intensity of site energy fluctuations.
Finally, the master equation for the density matrix $\rho$ (setting $\hbar=1$)
can be obtained adding both the non-Hermitian term and the dephasing term:
\be
\frac{d\rho_{hk}}{dt} =-i(\he\rho-\rho \he^\dagger)_{hk}-\gamma_\phi
(1-\delta_{hk})\rho_{hk}.
\label{mastereq}
\ee
Usually, in molecular aggregates, energy is measured 
in units of $\textrm{cm}^{-1}$, corresponding to energy divided by
$hc$. In these units, time is measured in $\textrm{cm}$ which
corresponds to  the mapping $t \to 2\pi c t$  ($ c \simeq 0.03 \
\textrm{cm/ps}$ is the speed of light) and  Eq.~(\ref{mastereq}) remains the same.
In the following all units of energy will be given in $\textrm{cm}^{-1}$ and in order 
to have time in $\textrm{ps}$ we need to divide it by  $ 2\pi c $.

\section{Exact solution for macroscopic variables}

Let us define the following macroscopic variables:
\be
p={\rm tr}\rho=\sum_{h=1}^N\rho_{hh}, \quad q=\sum_{h\ne k}\rho_{hk}.
\label{supvar}
\ee
As for their physical meaning, $p(t)$ describes the probability of finding the excitation in the ring (which will decay to zero since the system is open) and $q(t)$ is related to the presence of quantum coherences and it is a real number,
being the sum of all the off-diagonal elements of the Hermitian density matrix. 
Those variables completely characterize the radiative behavior of the system.

Following what is commonly done in literature 
in the case of interaction with the EMF~\cite{zhao},
we  give an analytical expression of the decay width  for a given state
$\rho$.
Since in our case the state $|S\rangle$, Eq.~(\ref{Sr}),
is the only decaying states, while the others do not decay at all,
 the decay width  of any state $\rho$
can be written as the probability 
 to be the
state $|S\rangle$  times its  decay width $N\gamma$,
\begin{equation}
\label{nad}
\Gamma_{\rho}= N\gamma
\langle S| \rho| S \rangle= \gamma \sum_{n,m} \rho_{nm}= \gamma (p + q).
\end{equation}
From the last expression it is clear that, if $p=1$ and $q=0$, we have the decay
width  $\Gamma_{\rho}=\gamma$, and that we need $q\ne 0$ to have Superradiance or Subradiance.
For instance, for the  superradiant state in 
Eq.~(\ref{Sr}), we have: $p=1,q=N-1$, so that $\Gamma_{\rho} =N
\gamma$. Since $|S\rangle$ is also 
a fully entangled state, this shows that $q$ represents the coherences responsible for
Superradiance.

The system of $N^2 \times N^2$ differential equations (\ref{mastereq})
decouples, so that we can write a close set of equations for the macroscopic variables:
\begin{equation}\label{system}
\left\{\begin{array}{ll}\dot p=-\gamma p-\gamma q,\\
\\
\dot q=-(N\gamma-\gamma)p-(N\gamma-\gamma+\gamma_\phi)q.
\end{array}\right.
\end{equation}
The dynamics of $p(t)$ and $q(t)$ can be obtained easily starting
from the  initial conditions
\be
\label{inicon}
p(0)=\sum_{h=1}^N\rho_{hh}(0) = 1 , \quad q(0)=q_0=\sum_{h\ne k}\rho_{hk}(0),
\ee
and it is given by
\begin{align}
p(t) &\mbox{}=-p_{_-}
e^{\lambda_+t}+p_{_{+}} e^{\lambda_-t},\label{pt2a}\\
q(t) &\mbox{}= p_{_-} \left(1+ \frac{\lambda_+}{\gamma}\right) e^{\lambda_+t}
- p_{_+}\left(1 + \frac{\lambda_-}{\gamma} \right) e^{\lambda_- t},\label{pt2b}
\end{align}
where
\be
\label{pt2o}
p_{_ \pm}   =\frac{\gamma+\lambda_{_\pm} + \gamma q_0}{\lambda_+-\lambda_-},
\ee
and  $\lambda_{\pm}$
are
the eigenvalues of the linear system (\ref{system}),
\be
\label{lambda}
\lambda_{\pm}=\frac{-N\gamma-\gamma_\phi\pm\sqrt{N^2\gamma^2+
\gamma_\phi^{2}+(2N-4)\gamma\gamma_\phi}}{2}.
\ee

Note that Eq.~(\ref{pt2a}) for the population $p(t)$ is in agreement
with the results obtained in Ref.~\cite{mukameldeph}, while the
results for coherences have not been discussed so far.
Note also that,  for $\gamma=0$, we have $\dot{p}=0$ and
$\dot{q}=-\gamma_{\phi}q$, so that coherences would simply decay
exponentially as $q(t) = q(0) \exp(-\gamma_{\phi}t)$. 
A very different situation arises
because of the coupling with a common decay channel.
It is interesting to study the limit for small and large $\gamma_\phi/N\gamma$. 
In the regime,  $\gamma_\phi/N\gamma\ll 1$, we have: 
\be
\label{lambda1}
\begin{array}{lll}
\lambda_+&=\displaystyle-\frac{\gamma_\phi}{N}+o\left(\frac{\gamma_\phi}{N\gamma}\right),\\ 
&\\
\lambda_-&=\displaystyle-N\gamma-\gamma_\phi\frac{N-1}{N}+o\left(\frac{\gamma_\phi}{N\gamma}\right).
\end{array}
\ee
This means that, in this regime, the dynamics are characterized by a cooperative
(proportional to the system size $N$) exponentially fast decay with rate $-N\gamma$ 
(superradiant), and a strongly suppressed decay with slope $\gamma_\phi/N$
(subradiant). We will call this the {\it superradiant regime}.

On the other hand, for
$N\gamma/\gamma_\phi \ll 1 $
one has 
\be
\begin{array}{lll}
\label{lambda1c}
\lambda_+&=-\gamma+o(N\gamma/\gamma_\phi),\\
&\\
  \lambda_-&=-\gamma_\phi[1+o(N\gamma/\gamma_\phi)].
\end{array}
\ee
In this regime any cooperative effect on the decay is lost and the long-time dynamics are dominated by the smallest exponent between $\gamma$ and $\gamma_\phi$.
This is the {\it non-superradiant regime}.

\section{Survival Probability}

The critical dephasing strength separating the
{\it superradiant regime} from the {\it non-superradiant regime}
can be studied by means of the survival probability, $p(t)$, 
for the excitation to be still in the system at time $t$. 
Note that,  for $\gamma_{\phi}=0$,
assuming as initial condition a superradiant state ($p_0=1,q_0=N-1$), Eq.~(\ref{Sr}), 
we have $p(t)=e^{-N\gamma t}$, while for any subradiant initial state
we have: $p(t)=p_0=1$.

\begin{figure}[t]
\centering
\includegraphics[width=0.42\textwidth]{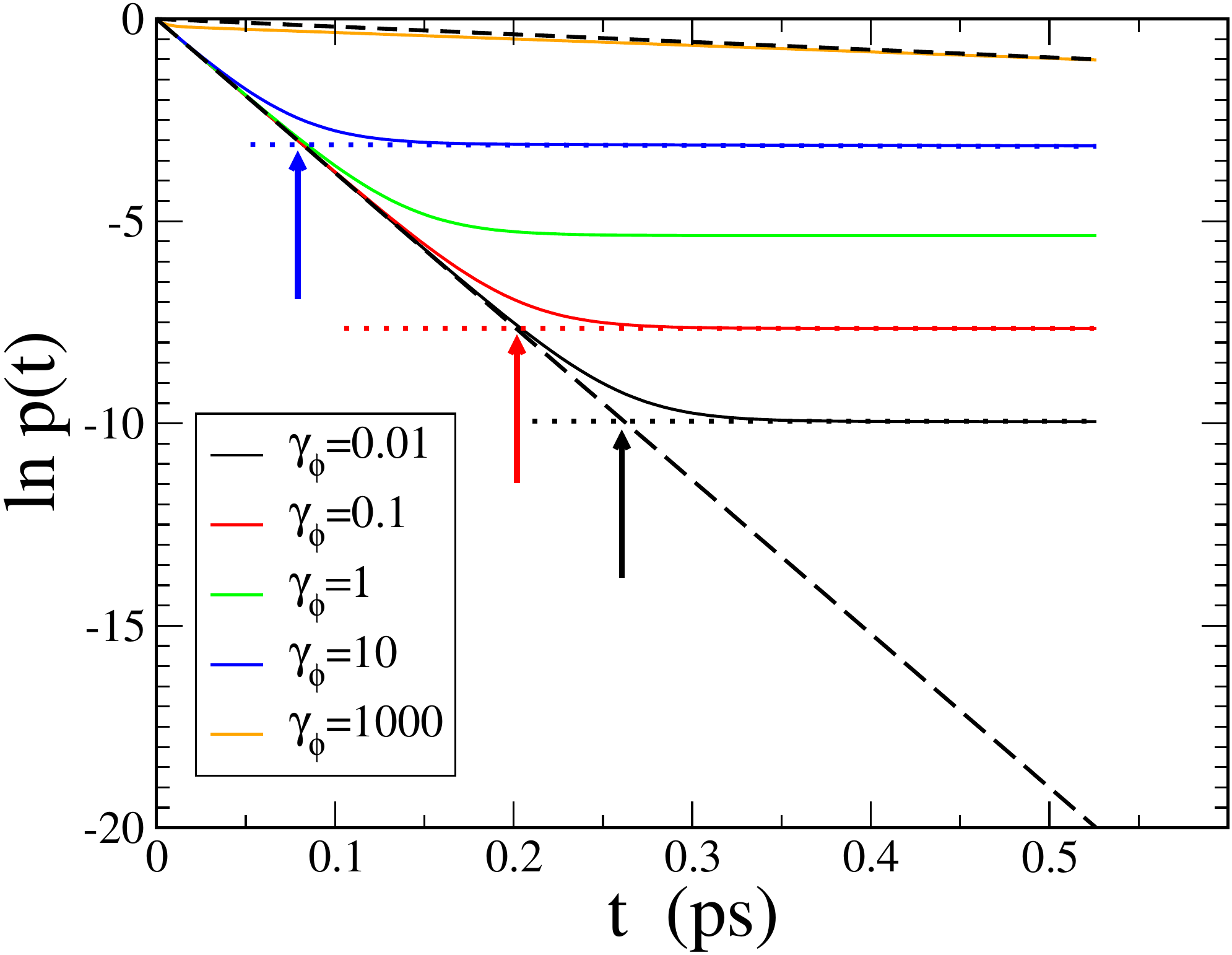}
\caption{(Color online) Survival probability, $p(t)$,   as a function of time, $t$,  
measured in $ps$, see discussion below Eq.~(\ref{mastereq}).
Different
 values of the dephasing rate $\gamma_{\phi}$ are shown as solid lines . 
The superradiant decay with rate $N\gamma$ and the decay with rate $\gamma$ are also shown
 as dashed lines. 
Parameters are $\Omega=1$,  $\gamma=10$ and $N=20$. The 
vertical arrows indicate the analytical expressions for the time $t^{*}$, 
Eq.~(\ref{tstarapprox}), at which the superradiant decay ends.
Dotted lines represent exponential decays with exponents $\gamma_\phi/N$.
Initial state is the superradiant extended state $|S\rangle$.
}
\label{fig:ring}
\end{figure}

In Fig.~\ref{fig:ring} we show the survival probability, $p(t)$,
 as a function of time  for different $\gamma_{\phi}$ values,
starting from the  superradiant state.
As one  can see, for $\gamma_\phi/N\gamma \ll 1 $, it
decreases  in time 
initially  as $e^{-N\gamma t}$
and, for $t> t^*$,  as $e^{-\gamma_\phi t/N}$ independently of $\gamma$.
It is relatively easy to estimate the transition time,  $t^*$, 
as the intersection, in logarithmic scale,  between the lines with  
slope $-N\gamma$
and $\gamma_\phi/N$, obtaining
\be
\label{tstarapprox}
t^{*}\approx -\frac{1}{N\gamma}\log\frac{\gamma_{\phi}}{N\gamma}.
\ee
The time $t^*$, which represents the timescale at which the 
superradiant transfer ends,  becoming subradiant,
has been shown
in Fig.~\ref{fig:ring}
by vertical arrows. 
On the other side,
for large enough dephasing rate $\gamma_{\phi} > N\gamma $,   
Superradiance is destroyed and the 
survival probability 
decays as  $e^{-\gamma t}$, see Fig.~\ref{fig:ring}. 

To determine the critical dephasing
strength at which the super and subradiant effect is destroyed,
we analyze the decay of both the superradiant state, Eq.~(\ref{Sr}), and of a 
particular subradiant state, 
\be
\label{anti}
\ket{A}=\frac{1}{\sqrt{N}}\sum_{i=1}^N (-1)^{i}\ket{i},
\ee
which is clearly  orthogonal to the superradiant one, ($\langle S|A\rangle = 0$).
Let us consider   the time, $\tau$, at which $p(\tau)=1/e$:
it is clear that Superradiance is completely destroyed when $\Lambda\equiv 1/\tau \simeq  \gamma$.
Note that $\Lambda/\gamma $ has also been called effective cooperation
number ($N_{\mathrm{eff}}$) in literature\cite{mukameldeph}.

\begin{figure}[t]
\includegraphics[width=0.45\textwidth]{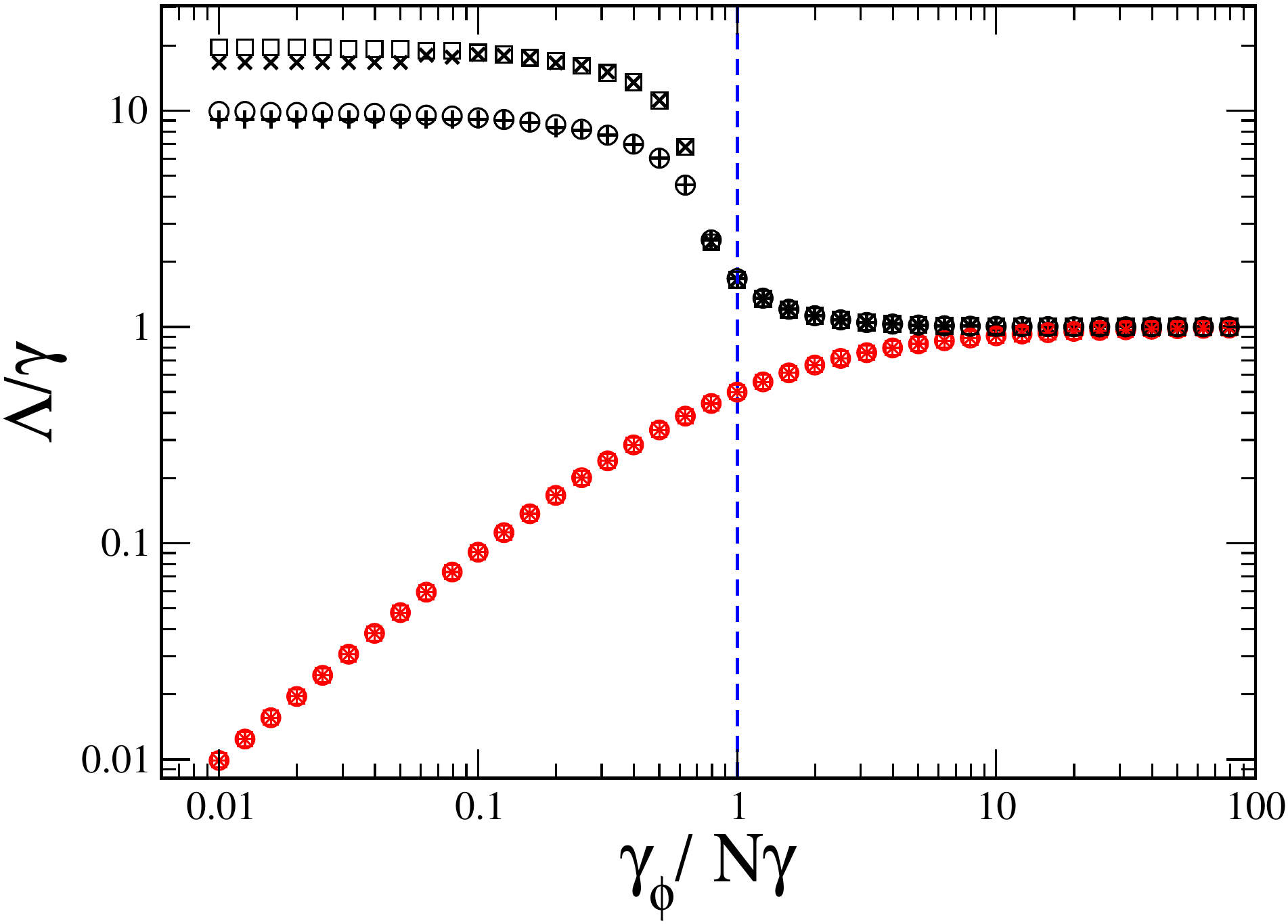}
\caption{ (Color online) $\Lambda/\gamma$ as a function of $\gamma_{\phi}/N\gamma$ for
an initial state :  extended, $|S\rangle$, (upper black symbols);
 antisymmetric    $|A\rangle$, (lower red  symbols).
Different symbols correspond to different parameters:  plus
($N=10, \Omega=1, \gamma=100$),   crosses
($N=20, \Omega=1, \gamma=100$),  circles
($N=10, \Omega=100, \gamma=1$),  squares
($N=20, \Omega=100, \gamma=1$).
 The vertical dashed  line is
the critical threshold $\gamma_\phi/N\gamma = 1$.} 
\label{ff2}
\end{figure}

As it is shown in Fig.~\ref{ff2}, the ratio $\Lambda/\gamma$, for the superradiant initial state, goes to $1$ when $\gamma_\phi/N\gamma \approx 1$. 
For the anti-symmetric initial state  Eq.~(\ref{anti}), we have that:
\begin{itemize}
\item for $\gamma_{\phi}/N\gamma < 1 $, 
$\Lambda$ $\propto \gamma_{\phi}/N$; 
\item for $\gamma_{\phi}/N\gamma > 1 $, the ratio $\Lambda/\gamma$ approaches $1$, in agreement 
with the analytical results presented above.
\end{itemize}
We also checked (not shown in Fig.~\ref{ff2})
that this behavior is
independent of $\gamma$ and shared by all of the subradiant initial states.

Fig.~\ref{ff2} also shows the robustness
of super and subradiant emergent effects to dynamical disorder: the amount of
dephasing needed to destroy the superradiant behavior increases with the
system size. Note that the same robustness was found in the presence of static
disorder~\cite{alberto, laltro}, so that it appears to be a general property of such coherent
emergent effects.
Another interesting result is that, at fixed dephasing rate, 
the decay rate of the subradiant states decreases
on increasing the system size (as $\gamma_{\phi}/N$).
This indicates that some kind of 
quantum coherence is cooperatively preserved in  presence of  Superradiance.
 
\section{Entanglement}
Here we would like to address the question whether other kinds of
quantum coherence, different from that needed to sustain Superradiance, 
are preserved by the coupling to a common decay channel.
Therefore, we will base our analysis  on a quantity 
used to quantify the global entanglement~\cite{photo2}:
\be
E[\rho]=-\sum_{i} \rho_{ii}\ln\rho_{ii}+ \sum_{i}\lambda_{i}\ln{\lambda_{i}},
\ee
where
$\lambda_i$ are the eigenvalues of the density matrix.
Note that $E[\rho]$ measures the entanglement between the many two-levels systems which
share the excitation.
Let us stress that, for an extended state, see Eq.~(\ref{Sr}),  
we have $E[\rho]= \ln(N)$, while, for a separable state, we
have $E[\rho]=0$.  Moreover, the von Neumann
entropy, $\sum_{i}\lambda_{i}\ln{\lambda_{i}} $,
vanishes for pure states, so that  
$E[\rho]=-\sum_{i} \rho_{ii}\ln\rho_{ii}$.

In the single exciton approximation it is also possible to compute the
concurrence~\cite{wootters} for any pair of two-levels systems $i\ne j$, which
is simply given by~\cite{photo2} 
$$C_{ij}=2|\rho_{ij}|.$$

\begin{figure}[t]
\centering
\includegraphics[width=0.47\textwidth]{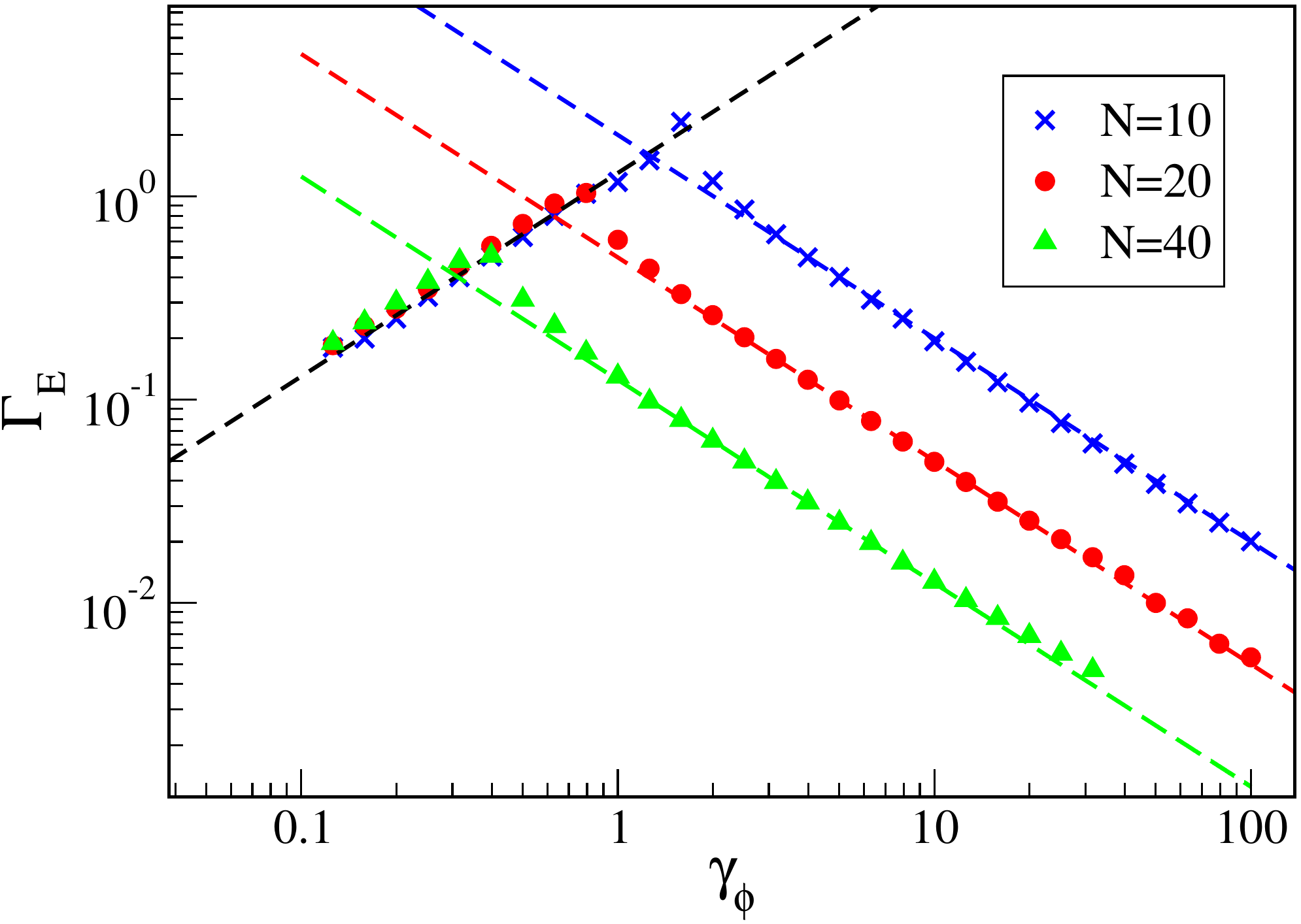}
\caption
{(Color online)
Dependence of the decay rate of the global entanglement 
for the closed system as a function
of the dephasing $\gamma_\phi$ for different $N$ values, as indicated in the legend.
Here is: $\gamma=0$, $\Omega=1$, and as initial state we chose the single
site $|k\rangle$. Dashed (black) line is the fitting $\Gamma_E = 1.3 \gamma_\phi$,
red, blue and green dashed lines stand for $\Gamma_ E = C(N)/\gamma_\phi$,
where $C(10) = 2, C(20) = 1/2, C(40) = 1/8$.
}
\label{g0}
\end{figure}

Before analyzing the effect of opening on the global entanglement,
let us discuss 
the case without  opening  ($\gamma = 0$).
Since the off-diagonal matrix elements are 
exponentially suppressed,
the steady state solution of Eq.~(\ref{mastereq}), is
given by,
$\rho_{ij} = \delta_{ij}/N$.
Thus, in our model of decoherence,  the density matrix becomes diagonal
with a homogeneous  distribution,
all the off-diagonal matrix elements and $E[\rho]$ (which represent the coherences)
vanish for large enough times. 
\begin{figure}[t]
\centering
\includegraphics[width=0.47\textwidth]{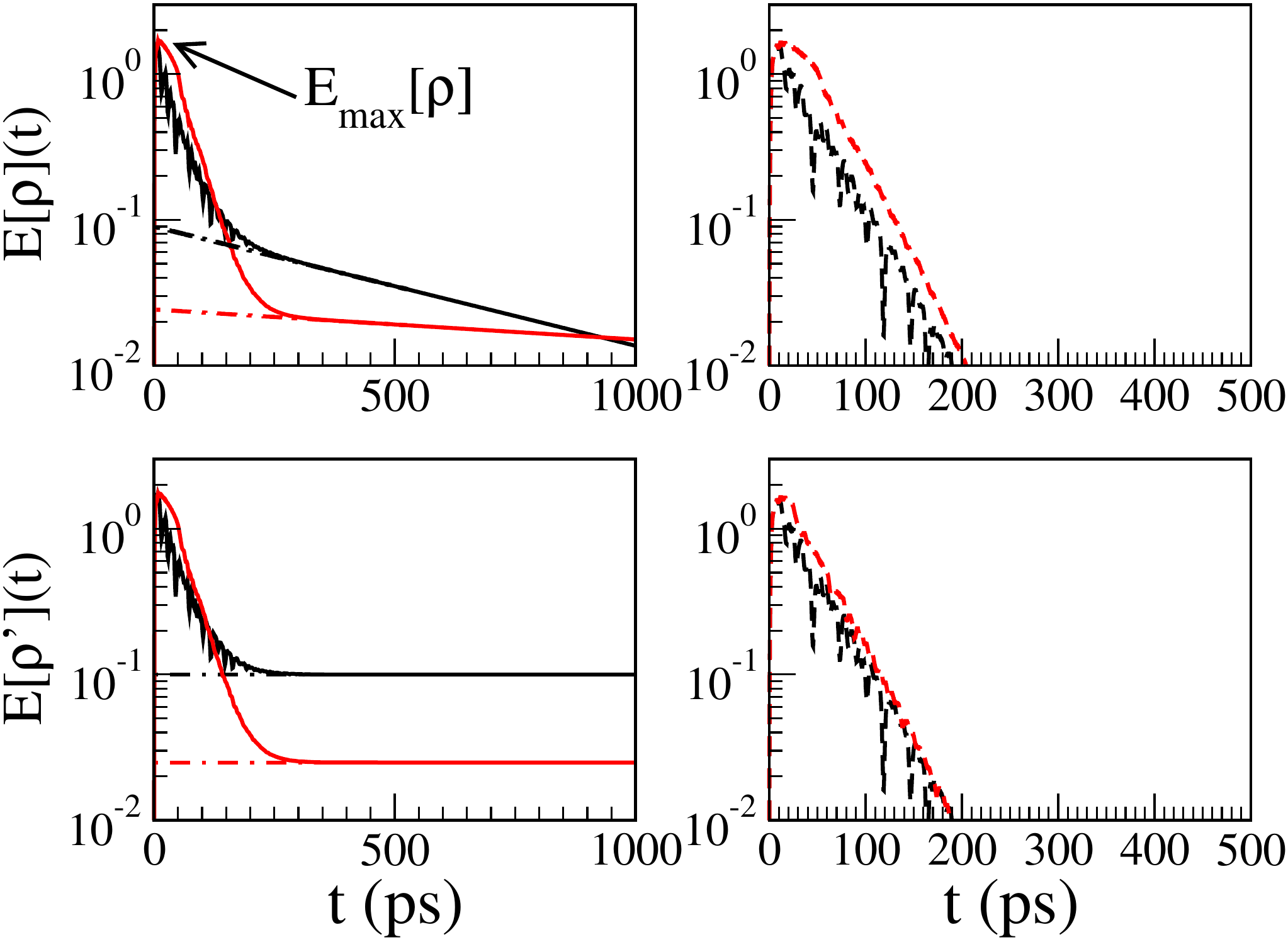}
\caption
{(Color online)
Upper panels: Global entanglement, $E[\rho]$, as function of time, $t$, starting
from a single site, $|i\rangle$,
for the open system with $\gamma=1$ (left panel) and
for the closed system with $\gamma=0$ (right  panel).
In the left panel the
analytical results for the asymptotic global entanglement are shown
as dot-dashed lines.
Different system sizes are considered: $N=10$ (black curves) and $N=40$ (red curves).
Lower panel: the same quantities 
for the normalized density matrix, $\rho'$, see text.
Data are $\Omega=1$ and $\gamma_{\phi}=0.1$.
}
\label{g1}
\end{figure}
Typically, one has that $E_\rho (t) \sim e^{-\Gamma_E t }$, for $t$ large.
The dependence of $\Gamma_E$ on $\gamma_\phi$ for the different parameters has been obtained 
by an exponential fitting in time and  is shown in Fig.~\ref{g0}.
As one can see,  $\Gamma_E$ is not monotone with $\gamma_\phi$, since
 it firstly increases linearly with $\gamma_\phi$ 
and then, for large dephasing, it decreases as $1/\gamma_\phi$. 
This fact
has been interpreted in literature as a manifestation
of the so-called quantum Zeno effect~\cite{lloyd}.
To summarize, in the case $\gamma=0$ we numerically found: 
\begin{equation}
\Gamma_E = \left\{
\begin{array}{lll}
1.3~ \gamma_\phi \quad &{\rm for } \quad &\gamma_\phi < 12.4~\Omega/N,\\
&\\
\displaystyle 
\frac{200~\Omega^2}{N^2\gamma_\phi} \quad &{\rm for } \quad &\gamma_\phi > 12.4~\Omega/N.
\end{array}
\right.
\label{gam0}
\end{equation}

Let us now consider the case $\gamma \ne 0$. 
Under very general assumptions, see Appendix A, it is possible
to obtain the asymptotic behavior (for $t \to \infty $) of the density matrix:
\be
\rho_{ij}(t) = e^{\lambda_+ t} \left[ a \delta_{ij} + (1-\delta_{ij}) b \right],
\label{rhoasym}
\ee
where
\be 
a = -\frac{p_-}{N}, \qquad b = \frac{p_-}{N(N-1)}\left(1+\frac{\lambda_+}{\gamma}\right),
\label{aandb}
\ee
and $ p_-, \lambda_+$ are defined in Eqs~(\ref{pt2o},\ref{lambda}).
Eq.~(\ref{rhoasym}) has been numerically verified
in a very large range of  parameters.

From the analytical expression, Eq.~(\ref{rhoasym}), it is easily deduced
that both global entanglement and concurrence decay exponentially in time, as
$\exp(\lambda_+ t) $.
In more detail, one gets the  asymptotic ($t\to\infty$) behavior for the  concurrence,
\begin{equation}
C_{ij} \sim 2 e^{\lambda_+ t} \left[
  \frac{p_-}{N(N-1)}\left(1+\frac{\lambda_+}{\gamma} \right) \right],
\label{C1}
\end{equation}
and for the global entanglement,
\be
\begin{array}{lll}
E[\rho] &\sim  \displaystyle \frac{-p_-}{N}  e^{\lambda_+ t}  \left \{ \left(2+
    \frac{\lambda_+}{\gamma} \right)
\ln \left(2+ \frac{\lambda_+}{\gamma}\right) + \right.\\
&\\
& \left. + \displaystyle \left(N-  2  - \frac{\lambda_+}{\gamma}\right) \ln \left[
1 - \frac{1}{N-1} \left(1+ \frac{\lambda_+}{\gamma}\right)
\right] \right \}.
\end{array}
\label{E1}
\ee
This means that, in the superradiant regime,
where $\lambda_+ = -\gamma_\phi/N$, see (\ref{lambda1}), the decay
of the entanglement is suppressed as we increase the system size.
On the other hand, for large dephasing, $\gamma_\phi \gg N\gamma$, where
Superradiance is destroyed, the entanglement decays as $\exp(-\gamma t)$, independently of the system size. Thus entanglement decay displays a
transition from a size-independent decay (for $\gamma_\phi \gg
N\gamma$) to a {\it cooperative sustained} decay (for $\gamma_\phi \ll
N\gamma$), where the entanglement decay becomes independent of $\gamma$ and
is suppressed on increasing the system size, $N$, 
 as $\gamma_{\phi}/N$. This proves that, in the superradiant regime, also other kinds 
of coherences are affected by cooperativity.

In Fig.~\ref{g1} we  analyze the behavior of 
$E[\rho]$, starting from a wave function localized on a single site,
$|\psi_0 \rangle = |i\rangle$. Since this state overlaps with the  
superradiant state with a small probability, $1/N$, it can be
considered mostly subradiant for large $N$. 
In Fig.~\ref{g1} (upper panels) we fix $\gamma_\phi=0.1$ and we
consider two different situations: with opening (left panel $\gamma=1$) and 
without opening (right panel $\gamma=0$),
both for different system sizes, $N=10,40$.
As one can see, the presence of the coupling to a common decay channel
produces the following effects:
\begin{enumerate}
\item it slows down the entanglement decay (compare the left panel with the right one);
\item it decreases the decay rate on increasing the system size, since 
the entanglement decays as $e^{-\gamma_{\phi} t /N}$ (compare the black with the red line
in the left panel).
\end{enumerate}


The asymptotic expansion, Eq.~(\ref{E1}), represents the entanglement at the time $t$ 
for a system that can decay  into the continuum. This means 
that the global entanglement decays not only because of the destruction of coherences, 
but also due to  the loss of probability into the continuum. 
It is interesting to consider what would be the global entanglement 
after a measurement which finds the excitation in the ring.
To this end, one should consider the density matrix $\rho'(t)= \rho(t)/Tr[\rho(t))]$ 
after the measurement given by the projection operator $\sum_i |i\rangle \langle i |$.
In this case one obtains a stationary entanglement
\be
\begin{array}{lll}
E[\rho'] &= \frac{1}{N} \left(2+ \frac{\lambda_+}{\gamma}\right)
\ln \left(2+ \frac{\lambda_+}{\gamma}\right) +\\
&\\
&+ \left(1-  \frac{2}{N}  - \frac{\lambda_+}{N\gamma}\right) \ln \left[
1 - \frac{1}{N-1} \left(1+ \frac{\lambda_+}{\gamma}\right)
\right],\\
\end{array}
\label{E2a}
\ee
(see Eq.~(\ref{E2}) in Appendix B),
 which is independent of time and initial
conditions. This finding is confirmed by our numerical results
for the entanglement of the normalized density
matrix, shown in Fig.~\ref{g1} (lower left panel).
Note that the entanglement of the normalized density matrix indicates the
entanglement present in the system at time $t$, after a measurement
which finds the excitation in the system.  
Note that while all the  coherences $\rho_{ij} \to 0 $, as $t\to \infty$,
the same does not happen for the normalized density matrix ${\rho'}_{ij}$.
The mathematical reason is that also $Tr(\rho) \to 0$   for $t \to \infty$.

\section{Maximal Entanglement}

\begin{figure}[t]
\centering
\includegraphics[width=0.45\textwidth]{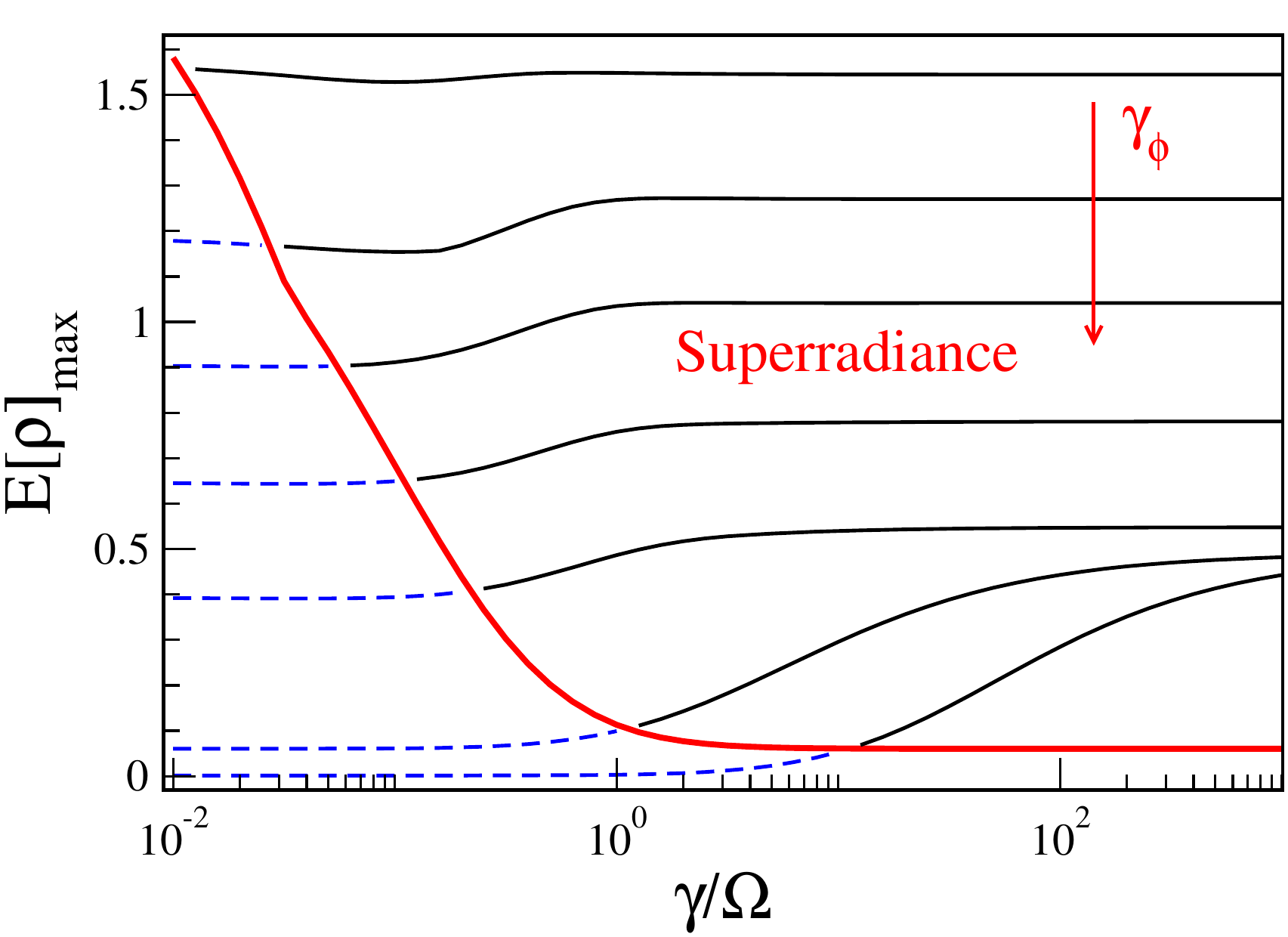}
\caption
{(Color online)
Maximal global entanglement $E[\rho]_{max}$  as a function of
the degree of opening $\gamma$ for fixed $N=10$ and $\Omega=1$.
Each curve has been obtained at fixed $\gamma_\phi=0.1, 0.25, 0.5, 1, 2, 10, 100$
(from the upper curve to the lower one), 
and it is shown as blue dashed in the regime characterized by $\gamma_\phi > N \gamma$ and
as full black  in the superradiant regime $\gamma_\phi < N \gamma$.
The thick red curve represents the superradiant transition $\gamma_\phi = N \gamma$, so
that the supperradiant region is above that curve.
As initial state we chose a single site $|i\rangle$.
}
\label{g2}
\end{figure}

Another interesting quantity to be studied is the maximal entanglement (value at the peak)
obtained during the whole evolution, see Fig.~\ref{g1}. 
In absence of coupling to the continuum, $\gamma=0$,  the coupling, $\Omega$,  between neighboring sites creates entanglement (with those initial conditions we have $E[\rho(0)] = 0 $), which is then exponentially suppressed in time by any chosen dephasing rate, $\gamma_\phi \ne 0$, see Fig.~\ref{g1}.
Maximal entanglement occurs at the  time
$\tau \simeq \hbar/\Delta E$, where $\Delta E\simeq 4\Omega/N$ is the unperturbed
energy spacing, and its maximal value decreases on increasing the
dephasing rate $\gamma_\phi$.
On the other hand,  the competing effects due to the coupling with the continuum of states,
$\gamma \ne 0$,  result in different behaviors depending on the choice of the parameters.
Results for maximal entanglement $E_{max}[\rho]$ are presented in Fig.~\ref{g2}.
As one can see, in the non-superradiant region, $N\gamma < \gamma_\phi$ (below the thick red line), $E_{max}[\rho]$ is almost independent of the coupling $\gamma$, while, on entering
the superradiant region, $N\gamma > \gamma_\phi $, there is a strong enhancement
of the maximal entanglement (mainly in the region of large dephasing), up to reach a saturation value.
In conclusion, we can say that, even in presence of a dephasing environment,
an optimal coupling to the continuum of states exists, such that
a maximal global entanglement is reached.

\section{Conclusions} 
We have analyzed a ring of two-levels systems in the limit where only
one excitation is present. We studied the combined effect of  different
environments on such a ring structure.
The two-levels systems are coupled to a
common decay channel,
which induces super and subradiant
behavior; moreover, they are also coupled to a dephasing environment,
modeled by a classical white noise (Haken-Strobl approach).
We have shown that coherent emergent properties, such as 
Superradiance and Subradiance, display a cooperative robustness to dephasing, in the sense that 
the critical dephasing rate needed to destroy these coherent effects increases with the system size. 
By analyzing the global
entanglement, we have demonstrated that the coupling to a common decay channel
is able to prevent the loss of coherences. Indeed, we have
shown that the entanglement decay is suppressed, 
in the superradiant regime, as we increase the system size. 
We have also derived an analytical expression for the asymptotic evolution of the
density matrix. 
In the future it will be interesting
to understand
how this asymptotic form depends on the details of the coupling to the
common decay channel.
This would allow to control the state
of a quantum system and its degree of entanglement by suitably coupling
it to one or more dissipative environments.

In order to compare our results with experimental data about molecular aggregates
and natural photosynthetic complexes one needs to estimate the dephasing rate.
Typically,
the superradiant regime described in this paper could be reached in realistic systems
such as J-aggregates, at low temperature or for a very large number 
of molecules in the aggregates. For instance in PIC Br J-aggregates we
have $(\hbar/\gamma)^{-1}  = 3.7\,\textrm{ns}$ and a 
dephasing rate of $10\,\textrm{ps}$ at
$1.5\,\textrm{K}$~\cite{spano,Jaggr}. Thus the superradiant regime
occurs for $N > 370$, which is easily reachable in J-aggregates.
On the other hand one should be careful to apply the results from Haken-Strobl
master equation  blindly, see discussion in the Introduction.

At room temperature the dephasing rate for  natural photosynthetic complexes,
such as LHI and LHII, is estimated to be on the order of $100 \ cm^{-1}$~\cite{expdeph, mukamelspano},
which is much larger than the superradiant decay rate. Indeed for the
LHI complex, we have $N=32$ and $\gamma \approx 10^{-3}\,\ cm^{-1}$~\cite{mukameldeph} for the coupling  w.r.t.
 the electromagnetic field, and  $\gamma \approx
10^{-2} \ cm^{-1}$~\cite{laltro} for the coupling w.r.t to reaction center.  
Thus the superradiant decay rate, $N \gamma $,
w.r.t. both environments is much smaller than the dephasing rate.
In this range of parameters the Haken-Strobl master equation predicts the total quenching of
Superradiance. On the other hand Superradiance w.r.t. the electromagnetic field 
has been found experimentally at room temperature in LHI and LHII complexes~\cite{vangrondelle}.
For J-aggregates~\cite{Jaggr} Superradiance has also been found in presence of a dephasing rate which
is one order of magnitude larger than the superradiant decay rate~\cite{spano}. 
These facts have been explained in Ref.~\cite{spano} by introducing a more sophisticated
model of the interaction with the phonon bath, showing that the Haken-Strobl approach
 overestimates the detrimental effect of dephasing.

In the future it would be interesting to 
verify the main results obtained here, which are the cooperative
robustness to disorder and the possibility to preserve coherences
through the coupling with a common decay channel, with more sophisticated
models of exciton-phonon interaction.

{\it Acknowledgments.} 
We would like to thank Giulio Giusteri, Robin Kaiser,
Mohan Sarovar, Rosario Fazio and Diego Ferrari for providing many useful discussions.


\appendix

\section{Asymptotic form of the density matrix}
Let us write the master equation 
Eq.~(\ref{mastereq}) in the form of a linear system
\be
\dot{\rho}_{hk} = \left( {\cal L } \rho\right)_{hk} = \sum_{nm}  {\cal L }_{hknm} \rho_{nm},
\label{lsys}
\ee
where the $N^2 \times N^2$ elements $ {\cal L }_{hknm} $ are time independent. 
It is easy to see that for large time, $t\to \infty$ ,  and  generic initial conditions the
asymptotic behavior of $\rho_{hk}(t)$ will be dominated by the smallest eigenvalue $\lambda_0 =
- \min_k |\lambda_k| $ of the four-indexes matrix ${\cal L } $, so that
\be
\rho_{hk} (t)  \sim  e^{\lambda_0 t}    {c}_{hk},
\label{lsys1}
\ee
where the coefficients ${c}_{hk}$ depend on the initial conditions and on the 
labels $(hk)$. 

Let us now assume a kind of ergodicity, in the sense that all diagonal and off-diagonal
elements  equal some real ($a$) and complex ($b+ic$) constant, respectively.
More specifically, we assume that the constants
$a,b$ might be dependent on the initial conditions, but not on the position $(hk)$.
We therefore conjecture the following asymptotic behavior:
\be
\rho_{ij}(t) \sim  e^{\lambda_0 t} \left[ a \delta_{ij} + (1-\delta_{ij}) (b+ic) \right].
\label{rhoasym1}
\ee
Since 
 the  asymptotic density matrix, Eq.~(\ref{rhoasym1}),
 should be a solution of the master equation, 
Eq.~(\ref{mastereq}), the following 
system for the real part should hold:
\begin{equation}\label{rsy}
\left\{\begin{array}{ll}
\lambda_0 a = -\gamma [a+ (N-1) b], \\
\\
a(\lambda_0+\gamma) = -\gamma(N-1)b.
\end{array}\right.
\end{equation}
Requiring the determinant to be zero, in order to have a non-null solution, we obtain
$$
\lambda_0 =\frac{-N\gamma-\gamma_\phi\pm\sqrt{N^2\gamma^2+
\gamma_\phi^{2}+(2N-4)\gamma\gamma_\phi}}{2}.
$$
Since  $|\lambda_+| < |\lambda_-|$, one gets $\lambda_0 = \lambda_+$.
In the same way, it is easy to show that the imaginary part of the system can be solved
only by $c=0$.

Using Eqs.~(\ref{pt2a}) and (\ref{pt2b}), and taking into account that 
$|\lambda_+| < |\lambda_-|$, one gets immediately
Eq.~(\ref{rhoasym}):
\be
\label{asex}
\rho_{ij}(t) \sim  e^{\lambda_+ t} \left[ -\frac{p_-}{N} \delta_{ij} + 
\frac{(1-\delta_{ij})p_-}{N(N-1)}\left(1+\frac{\lambda_+}{\gamma}\right)
 \right].
\ee

\section{Asymptotic expression for renormalized density matrix  and entanglement}

From the asymptotic form of the density matrix, Eq.~(\ref{rhoasym}),
we easily get the asymptotic form of $\rho'= \rho/Tr(\rho)$:
\be
\label{asexp}
\rho_{ij}(t)' \sim   \left[ \frac{1}{N} \delta_{ij} + (1-\delta_{ij})  
\frac{1}{N(N-1)}\left(1+\frac{\lambda_+}{\gamma}\right)
 \right].
\ee
To compute the global entanglement, one has to compute the von Neumann entropy,
namely the eigenvalues of such a matrix, which are given by:
\begin{equation}\label{eige}
\left\{\begin{array}{ll}
\xi_1 = \displaystyle \frac{1}{N}\left(2+ \frac{\lambda_+}{\gamma}\right),\\
\\
\xi_k =  \displaystyle \frac{1}{N}\left[ 1 -  \frac{1}{N-1}\left(1+ \frac{\lambda_+}{\gamma}\right)\right],
\quad {\rm for}, \quad k=2,...,N.
\end{array}\right.
\end{equation}
Simple algebra then gives
\be
\begin{array}{lll}
E[\rho'] &= \frac{1}{N} \left(2+ \frac{\lambda_+}{\gamma}\right) 
\ln \left(2+ \frac{\lambda_+}{\gamma}\right) +\\
&\\
&+ \left(1-  \frac{2}{N}  - \frac{\lambda_+}{N\gamma}\right) \ln \left[
1 - \frac{1}{N-1} \left(1+ \frac{\lambda_+}{\gamma}\right)
\right].\\
\end{array}
\label{E2}
\ee
In the limit of large $N$, one easily finds
\be
\label{glen}
E[\rho'](t) \sim \frac{1}{N }\left( 2\ln2 -1 \right)+ o(1/N^2),
\ee
where terms of order $1/N^2$ have been neglected.
\\
\par
Finally, for the asymptotic behavior of concurrence we get
\begin{equation}
C_{ij} = 2 |\rho'_{ij}| \sim 2  \left[
  \frac{1}{N(N-1)}\left(1+\frac{\lambda_+}{\gamma} \right) \right].
\label{C2}
\end{equation}

\end{document}